\documentclass[a4paper]{aa}
\usepackage{graphics,psfig,times}
\begin{document}

\title{Evaporation and condensation of spherical interstellar clouds.
       Self-consistent models with saturated heat conduction and cooling}

% \subtitle{}

\author{W. Vieser \inst{1,2}
\and
        G. Hensler \inst{1}
}

\offprints{G. Hensler,\\ \email{hensler@astro.univie.ac.at}}

\institute{Institute of Astronomy, University of Vienna, T\"urkenschanzstr.\ 17,
           A--1180 Vienna, Austria 
\and
           Christoph-Probst-Gymnasium, Talhofstr.\ 7, D--82205 Gilching, Germany
}

\date{Received 02 August 2005/ Accepted 17 August 2007}

% \abstract{}{}{}{}{} 
% 5 {} token are mandatory
 
  \abstract
{Shortened version:} 
{The fate of interstellar clouds embedded in a hot tenuous medium depends on
whether the clouds suffer from evaporation or whether material condensates 
onto them. Analytical solutions for the rate of evaporative mass loss 
from an isolated spherical cloud embedded in a hot tenuous gas are deduced 
by Cowie \& McKee (1977). In order to test the validity of the analytical 
results for more realistic interstellar conditions the full hydrodynamical
equations must be treated.}
{Therefore, 2D numerical simulations of the evolution of interstellar clouds 
are performed with different internal density structures and surrounded 
by a hot plasma reservoir. Self-gravity, interstellar heating and cooling
effects and heat conduction by electrons are added. 
Classical thermal conductivity of a fully ionized hydrogen plasma 
and saturated heat flux are considered.}
{Using pure hydrodynamics and classical heat flux we can reproduce the 
analytical results. Heat flux saturation reduces the evaporation rate by 
one order of magnitude below the analytical value, because the density 
distribution changes drastically during the simulation. 
This main result still holds with self gravity or different cloud density 
structures while keeping the cloud radius and surface temperature constant. 
The evolution changes, however, totally for more realistic conditions 
when interstellar heating and cooling effects stabilize the self-gravity. 
The evaporation then turns into condensation, because the additional 
energy by heat conduction can be transported away from the interface 
and radiated off efficiently from the cloud's inner parts.}
  % conclusions heading (optional)
{The consideration of a saturated heat flux is inevitable for interstellar 
clouds embedded in hot tenuous gas.
Under more realistical conditions with radiative cooling heat conduction 
leads to condensation in contrast to analytical predictions which
require evaporation. This provides an efficient way to accrete and 
mix intercloud material into clouds.}

\titlerunning{Evaporation and condensation of spherical interstellar clouds}

\maketitle

%________________________________________________________________

\section{Introduction}

The Interstellar Medium (ISM) can be described as an inhomogeneous ensemble of
three phases (McKee \& Ostriker \cite{mo77}). 
The cold neutral phase with temperature $T \sim 80 \mbox{ K}$ and 
density $n \sim 40 \mbox{ cm}^{-3}$ is represented by the cores of
molecular clouds which are confined by a warm neutral to slightly ionized 
medium ($T \sim 8000 \mbox{ K}$, $n \sim 0.3 \mbox{ cm}^{-3}$). 
These two components
are in pressure equilibrium if the gas is externally heated and can cool
radiatively (Field et al. \cite{f69}). According to McKee \& Ostriker they are 
embedded in a third phase of the ISM:
the hot dilute intercloud medium (HIM) with $T \sim 10^6 \mbox{ K}$ and
$n \sim 10^{-3} - 10^{-4} \mbox{ cm}^{-3}$. Since this
component is produced locally by supernova explosions at even higher 
temperatures and densities, it can originally not
be in pressure equilibrium with the cooler phases and has therefore to
expand vehemently. During this expansion  
shocks arise and the HIM penetrates through the ambient clumpy
ISM. Denser clouds cannot be swept-up but are passed by the HIM so that
they become embedded therein. Due to the strong discrepancy of the
physical states between the phases an interface has to form where the 
hot phase and the molecular cloud are in contact. Temperatures and densities
are connected through steep gradients that lead to energy and mass transfer.
At this interfaces heat conduction plays an important role and is of prime
importance whether the clouds evaporate or condense and therefore grow in 
mass. Analytical approximations require that small clouds in a hot plasma
suffer from evaporation and will therefore be only short lived. The fact
that interstellar clouds embedded in a hot medium are observed 
on the other hand suggests, that they are stabilized against
evaporation so that their destruction is delayed. 

These observations include {\it{High-Velocity Clouds}} (HVCs), H{\sc{i}} 
entities consisting of a multi-phase structure (Wakker \& Schwarz \cite{ws91}) 
and characterized by radial velocities that are
incompatible with simple models of the differential rotation of the galactic
disk (see Wakker \& van Woerden (\cite{ww97}) for a recent review). 
Distance measurements of the cloud complexes remain difficult.
For at least two of them upper limits for their distances are deduced by
Danly et al. (\cite{d93}), Keenan et al. (\cite{k95}) and 
van Woerden et al. (\cite{w97}) which locate them
in the hot galactic halo. Therefore, interactions with the hot
rarefied halo gas are expected. Indications for such interactions are
the detection of 
morphological perculiarities like head-tail structures (\cite{b00};
\cite{b01}). Interactions with gas belonging to the galactic disk and
having lower velocities are deduced from the observation of 
so-called {\it{velocity bridges}} (\cite{pi96}), fusions of 
high- and low-velocity gas in the velocity-position space.

Further examples for interstellar clouds in a stream of hot gas are
cometary globules, like G134.6+1.4, which is located in
a galactic chimney (\cite{no96}). This chimney is associated with the 
{H{\sc ii}} region W4 and forms a cone in the H{\sc i} layer above W4 
with a diameter of about 100 pc. 
At the bottom of the chimney the starcluster IC 1805 with 9
O stars can be found (\cite{he96}; \cite{ta99}). An intense
interaction between the shocked hot stellar winds and G134.6+1.4 is
expected. Energy exchange between the cold molecular to neutral gas of
the globule and the hot chimney gas leads either to an 
evaporation and therefore a destruction of the cloud what is not observed
or to condensation of hot gas onto the globule and therefore to a
stabilization. 

Previous papers dealing with the evaporation and condensation of molecular
clouds only solve the time-independent energy equation without taking
temporal effects into account like mass changes, heating, etc., and 
dynamical effects like mixing of warm cloud material with
the hot plasma and surface instabilities. In this paper we examine the
evolution of interstellar clouds 
that are embedded in a hot, dilute medium by solving the full hydrodynamical
equations numerically and furthermore following the heat conduction
process in detail. In \S2 we discuss the predicted evolution that
is expected refering to the papers of Cowie \& McKee 
(\cite{cm77}, hereafter: CM77) and
Dalton \& Balbus (\cite{db93}, hereafter: DB93).
The treatment of heat conduction in the context of hydrodynamical simulations 
is described in \S3. In \S4 we introduce the different models and show the
results of the simulations. 
Conclusions are drawn in \S5.

\section{Heat Conduction}

The energy excange between the phases as a consequence of heat 
conduction is described by a heat flux $\vec{q}$. In the classical case 
this flux is calculated assuming a diffusion appoximation leading to the 
formula given by Spitzer (\cite{sp62}):

\begin{equation}{\label{eq201}}
\vec{q}_{\rm{class}} = - \kappa \cdot \vec{\nabla} T
\end{equation}

with the heat conduction coefficient

\begin{equation}{\label{eq202}}
\kappa = \frac{1.84 \times 10^{-5} T^{5/2}}{\ln \Psi}
\mbox{ erg s$^{-1}$ K$^{-1}$ cm$^{-1}$} \quad ,
\end{equation}

where the Coulomb logarithm is
 
\begin{equation}{\label{eq203}}
\ln \Psi = 29.7 + \ln \left [ \frac{T_{6,e}}
{\sqrt{n_e} } \right ] \quad ,
\end{equation}

with the electron density $n_e$ and the electron temperature 
$T_{6,e}$ in units of $10^6$K. 

This description breaks down if the local temperature scaleheight falls 
below the mean free path of the conducting electrons. In this case the 
heat flux is replaced by a flux-limited form the so-called saturated 
heat flux (CM77):

\begin{equation}{\label{eq204}}
|\vec{q}_{\rm{sat}}| = 5 \Phi_s \rho c^3
\end{equation}

with the sound speed $c$ and density $\rho$. $\Phi_s$ is an efficiency 
factor less than or of the order of unity, which embodies some
uncertainties connected with the flux-limited treatment 
and flux suppression due to magnetic fields.
This method yields results in good agreement with 
laser-fusion experiments (\cite{mn73}; \cite{mk75}) and numerical
simulations solving the Fokker-Planck equation (\cite{kr81};
\cite{mv82}; \cite{c84}).

The heat conduction coefficient $\kappa$ is a function of the ionization
state of the medium. At temperatures lower than $10^4$K the ionization
fraction of the gas is negligible and with this the heat conduction by
electrons. Inside the cloud only conduction by neutrals may play a role for
the temperature distribution. On the other hand the temperature gradient
in this part of the cloud is too smooth to create a significant heat flux 
so that heat conduction becomes negligible.  
The evaporation and condensation of interstellar clouds was first investigated
by Zel'dovich \& Pikel'ner (\cite{zp69}) and later by Penston \& Brown
(\cite{pb70}) in a plane parallel approximation. The more realistic case
of spherical clouds was examined by Graham \& Langer (\cite{gl73}). They
solved the time-independent energy equation including heat
conduction. As a main
result they concluded that clouds with radii below a critical radius suffer
from evaporation while clouds with larger radii gain mass by condensation of
material onto the cloud surface because radiative cooling exceeds the heat
input due to heat conduction. Numerical simulations were performed by
Chevalier (\cite{c75}) in order to describe the evolution of interstellar
clouds in young supernova remnants. Although the temperatures rose up to
about $10^7$ K he took only the classical heat flux into account and, by this,
overestimated the mass-loss rate by some orders of magnitude.

\begin{figure}[h]
\psfig{figure=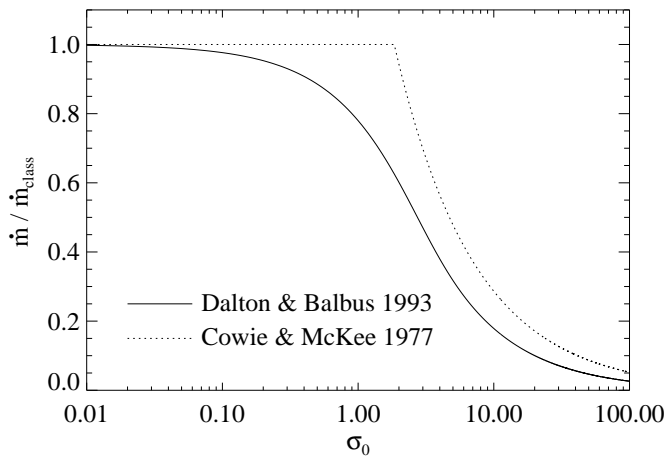,width=8.8cm,angle=0}
\caption{\label{fig01}Mass-loss rate $\dot{m}$ normalized to the
classical value $\dot{m}_{\rm{class}}$ (eq.~\ref{eq207}) as a function
of the global saturation parameter $\sigma_0$. Comparison of the rates
refering to CM77 (dotted) and DB93 (solid curve).}
\end{figure}

CM77 made a general study of the evaporation of spherical clouds including 
saturation effects. In the treatment of the saturation of nonmagnetic,
nonradiative, spherical evaporation, the domain outside the cloud
is broken up into three zones, each of which is treated seperately. These
zones are joined by prescribed boundary conditions for the heat flux and
the temperature. The heat conduction of the innermost zone which inner 
boundary is the cloud surface is described by the classical heat
flux. For larger radii the temperature rises and the heat flux
becomes saturated. CM77 treated this by abruptly changing the mathematical
form of the conductivity at a well-defined saturation radius. The temperature
rises steeply through the saturation zone until it reaches the value of the
ambient gas. At this radius, CM77 switched back to the classical description
of the heat flux. Beside a local saturation parameter $\sigma$ as the ratio
of the classical to the saturated heat flux,

\begin{equation}{\label{eq205}}
\sigma = \left| \frac{\kappa}{5 \Phi_s \rho c^3} \frac{dT}{dr} \right| 
\end{equation}

they introduced a global saturation parameter $\sigma_0$ 

\begin{equation}{\label{eq206}}
\sigma_0 = \frac{2 \kappa_f T_f}{25 \Phi_s \rho c^3 R} =
\left( \frac{T_f}{1.54 \times 10^7 \rm{K}} \right)^2 \frac{1}{n_f
\Phi_s R_{\rm{pc}}} 
\end{equation}

in order to ascertain the influence of saturation effects only by knowing 
the cloud radius $R$ or $R_{\rm{pc}}$ the same in parsec and the heat 
conduction coefficient, the temperature and particle density of the ambient 
medium $\kappa_f$, $T_f$ and $n_f$. For $\sigma_0 \le 1$ the mass-loss is
given by

\begin{equation}{\label{eq207}}
\dot{m}_{\rm{class}} = \frac{16 \pi \mu \kappa_f R}{25 \rm{k}}
\end{equation}

where k is Boltzmann's constant and $\mu$ is the mean molecular weight of 
the ambient plasma. For larger values of $\sigma_0$ the mass loss is a
decreasing function of $\sigma_0$ (see Fig.~\ref{fig01}).

In a following paper McKee \& Cowie (\cite{mc77}) analysed the influence
of radiative cooling on the evaporation rate. They found out that the 
additional heat input due to heat conduction is compensated by
radiative cooling for $\sigma_0 < 0.027/\Phi_s$. Also in this study 
the authors devided the cloud environment into three zones where the 
physics are described in different ways. In the inner and the outer
zone radiative
losses are larger than the spherical divergence of the heat flux, while in
the zone between the opposite is true and radiative losses are neglegted.
Because of the fact that only the classical heat flux is considered, this
approach can be only considered as a very crude estimation for the mass loss
rate. Saturation reduces the energy input into the cloud so that
radiation cooling can exceed this additional heat input.

Begelman \& McKee (\cite{bm90}) investigated the evolution of a two-phase
medium consisting of clouds embedded in a hot plasma. They showed that 
heating and cooling processes dominate the hot phase and heat conduction can
be neglected if the length scales of the
relevant structures are larger than a critical lenght, the so-called 
Field length:

\begin{equation}{\label{eq208}}
\lambda_{{\rm{F}}} \equiv \left[ \frac{\kappa T}{n^2 {\cal{L}}_M}
\right]^{1/2}
\end{equation}

where ${\cal{L}}_M$ defines the maximum heating or cooling rate.
The Field length is the maximum length scale on which energy transport
by heat conduction is effective. So the temperature distribution of 
structures with $r \ll \lambda_{{\rm{F}}}$ that are embedded in the hot 
plasma is dominated by heat conduction while this process can be neglected
for $r \gg \lambda_{{\rm{F}}}$. For an isolated cloud they concluded that
the classical mass-loss rate is valid for clouds with 
$R \ll \lambda_{{\rm{F}}}$ while condensation occurs for 
$R \approx 0.24 - 0.36 \lambda_{{\rm{F}}}$.

DB93 recovered the considerations of CM77 but this time used an
effective heat flux that is the harmonic mean of the
classical and the saturated flux in order to get a smooth transition
between both regimes:

\begin{equation}{\label{eq209}}
q_{\rm{eff}} = - \frac{\kappa}{ 1 + \sigma } \frac{dT}{dr} 
\end{equation}

\begin{figure}[h]
\psfig{figure=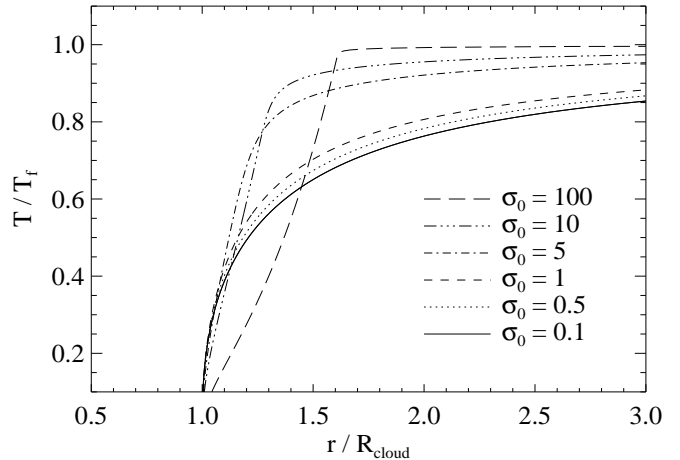,width=8.8cm,angle=0}
\caption{\label{fig02}Temperature profiles for different values of 
$\sigma_0$ derived from DB93.}
\end{figure}

Also in this description the time-independent energy equation can be solved
analytically. This leads to the temperature distribution outside the cloud
and also to the mass-loss rate as a function of $\sigma_0$. The mass loss 
rate is compared with the one derived by CM77 in Fig.~\ref{fig01}. The 
temperature distributions calculated by the formulae of DB93 for
different values of $\sigma_0$ (Fig.~\ref{fig02}) reflect the two
different mathematical descriptions of the heat flux. The solutions
for the classical flux ($\sigma_0 \le 1$) show a steep temperature
increase at the edge of the cloud and a decreasing gradient for larger
$r$. For large $\sigma_0$ the
curvature is positive near the cloud surface so that $T_f$
is reached at smaller $r$ than in the classical case.    

Because only the energy equation is solved in CM77 as well as in DB93, 
dynamical effects like mixing of the cold cloud material with the hot
plasma are not considered. On the other hand, it must be expected that the 
cloud material that is detached from the cloud surface is still cooler 
than the ambient plasma because only a slight rise in temperature of
the cloud material leads to its overpressure that decouples 
the material from the cloud's surface. This will influence
the temperature distribution near the surface and, by this, alter
the mass-loss rate accordingly. Therefore, the hydrodynamical
treatment is essential in order to explore to what extent the mass
flow by evaporation and condensation and further gas-phase mixing changes
the physical conditions around the interface.

%----------------------------------------------
%----------------------------------------------

\begin{table*}[t]
\centering
\caption{\label{t1}Model parameters and physical processes considered
in the various simulations. For all models: $R_{\rm{cloud}} =$ 41 pc,
$T_f = 5.6 \times 10^6$ K, $n_f = 6.6 \times 10^{-4}$ cm$^{-1}$,
$\sigma_0 =$ 4.88.}
\begin{tabular}{|l|c|c|c|c|c|c|c|c|c}
\hline
Model
& numerical grid  
& resolution   
& density \& temperature 
& \multicolumn{2}{c|}{heat conduction}
& self-
& heating/ 
& mass \\ \cline{5-6}   
& (cell$^2$)
& (pc cell$^{-1}$) 
& profile
& class.
& class. \& sat. 
& gravity
& cooling 
& (M$_{\odot}$) \\ \hline \hline 
R1a & 800 $\times$ 400 & 1 & homogen. & + & - & - & - & $3.4 \times 10^4$ \\
R1b & 800 $\times$ 400 & $0.5$ & homogen. & + & - & - & - & $3.4 \times 10^4$\\
R2a & 200 $\times$ 100 & 1 & homogen. & - & + & - & - & $3.4 \times 10^4$\\
R2b & 1000 $\times$ 500 & 1 & homogen. & - & + & - & - & $3.4 \times 10^4$\\
R2c & 900 $\times$ 450 & $0.33$ & homogen. & - & + & - & - & $3.4 \times 10^4$\\
R3  & 200 $\times$ 100 & 1 & core-halo & - & + & - & - & $6.4 \times 10^4$\\
R4  & 200 $\times$ 100 & 1 & core-halo & - & + & + & - & $6.4 \times 10^4$\\
R5  & 200 $\times$ 100 & 1 & core-halo & - & + & + & + & $6.4 \times 10^4$\\
\hline 
\end{tabular}
\end{table*}

%---------------------------------------------
\section{Hydrodynamical Treatment}

The spherical geometry of the cloud suggests to use a one dimensional 
spherical symmetric hydro-code. With a limitation to only one spatial 
dimension it is impossible to follow the growth of instabilities that 
are expected when cold dense gas is accelerated into warm rarefied gas. 
Therefore the evolution of clouds in a hot plasma is studied
by two-dimensional hydrodynamical simulations. The Eulerian equations 
are solved on a rectangular cylindrically symmetric 
``staggered grid'' which is of second order in space and based on the 
prescription of Rozyczka (1985). This code was extensively tested and 
used by different authors (e.g. Yorke \& Welz 1996).
The grid parameters, the resolution, the density/temperature profile of
the initial cloud model and the considered physical processes are
listed in Table~\ref{t1} for the different models. 

The different setups for the homogeneous clouds are chosen in order to
test the boundary conditions (model R2b) and the influence of
resolution on the mass-loss rate (model R1b, R2c). 
  
The boundary conditions at the upper, the left-hand and the right-hand
sides are semi-permeable to allow for an outflow of gas from the 
computational domain. The physical parameters at the lower boundary,
the symmetry axis, are mirrored. In order to trace the condensation of
the hot plasma onto the cloud a new quantity ``colour'' is introduced 
that is set in each cell to the density fraction of hot
ISM. At the beginning only the cells around the cloud possess a 
non-zero ``colour'' of value unity. During the
calculation this quantity is advected like the others as e.g.
mass or energy density. 
For the advection we use the monotonic transport (van Leer \cite{vl77}). 
The Poisson equation for self-gravity was solved for the corresponding 
model at each tenth timestep because significant density changes 
of the cloud structure happen on a much larger timescale than the 
dynamical one. The energy equation includes 
heating, cooling and heat conduction, according to what physical
process is included in the model:

\begin{equation}{\label{eq301}}
\frac{\partial e}{\partial t} +
\vec{\nabla} \cdot (e \vec{v}) =
-P \vec{\nabla} \cdot \vec{v} + \Gamma - \Lambda - \vec{\nabla} \cdot 
\vec{q} \quad .
\end{equation}

Here $e$ denotes the energy density, $\vec{v}$ the velocity, $P$ the pressure,
$\Gamma$ the heating rate, $\Lambda$ the cooling rate and $\vec{q}$ the
heat flux. The equation of state for an ideal gas is assumed to be valid:

\begin{equation}{\label{eq302}}
P=(\gamma - 1)e \qquad \mbox{with} \qquad \gamma = 5/3
\end{equation}

The used cooling function assumes collisional ionisation equilibrium and
is a combination of the function introduced by B\"ohringer \& Hensler
(\cite{bh89}) for $T>10^4$ K and solar metallicity and by 
Dalgarno \& McCray (\cite{dm72}) for the lower temperature regime.
The heating function considers cosmic rays (Black \cite{b87}),
X-rays and the photoelectric effect on dust grains 
(de Jong \cite{dj77}; de Jong et al. \cite{dj80}).
This detailed description of heating and cooling processes cannot 
be incorporated in analytical work like in CM77 that approximates the 
cooling function by four power-law segments. For temperatures 
above $10^4$ K we used a more up-to-date description than CM77 
who derived the cooling function in this temperature regime from 
Raymond, Cox \& Smith (\cite{rcs76}). Nevertheless, the cooling function 
used in our simulations is similar to the one used by CM77. 

The heat flux is calculated by taking both, the classical and the
saturated flux into account. In order to apply a smooth
transition between classical and saturated regime we use the
analytical form by Slavin \& Cox (\cite{sc92})

\begin{equation}{\label{eq303}}
\vec{q}=|\vec{q}_{\mbox{\tiny sat}}| \left(
1- \exp \left[ -\frac{\vec{q}_{\mbox{\tiny class}}}
                     {|\vec{q}_{\mbox{\tiny sat}}|}
        \right]
	\right) \; .
\end{equation}

This guarantees that the smaller flux is taken if both differ 
significantly. 
The heat flux due to electron diffusion is calculated separately using an
implicit method which follows a scheme introduced by 
Crank \& Nicolson (\cite{cn47}) 
and Juncosa \& Young (\cite{jy71}). To couple the two directions in space the 
method of fractional steps by Yanenko (\cite{y71}) is used.
A detailed description of the implementation as well as a numerical
test of the heat conduction code can be found in the online appendix.

The simulated clouds are embedded in a hot plasma with a given temperature 
$T_{\rm{f}} = 5.6 \times 10^6$ K and
particle density $n_{\rm{f}} = 6.6 \times 10^{-4}$ g cm$^{-3}$.
The most realistic cloud model numbered as R5 is generated 
for hydrostatic and thermal equilibrium under the constraint 
of spherical symmetry:

\begin{equation}{\label{eq304}}
\rho \, \vec{\nabla} \cdot \Phi = - \vec{\nabla} \cdot P
\end{equation}

\begin{equation}{\label{eq305}}
\Gamma(\vec{r}) = \Lambda(\vec{r})
\end{equation}

\begin{figure}[h]
\psfig{figure=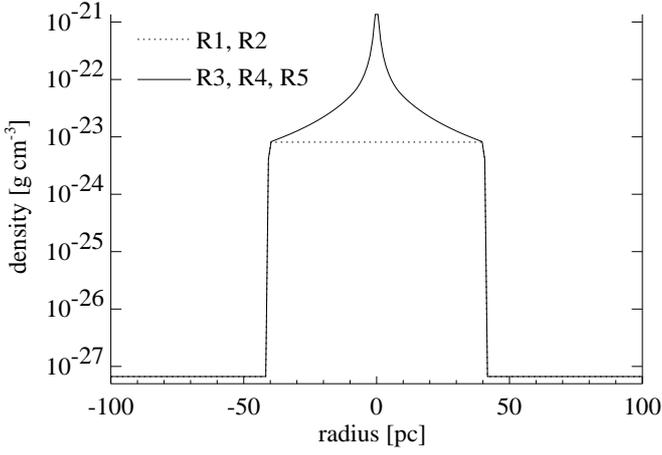,width=8.8cm,angle=0}
\caption{\label{fig03}Density profiles for the models with homogeneous
density distribution (dotted) and for those with pronounced central
peak (solid line).}
\end{figure}

\begin{figure}[h]
\psfig{figure=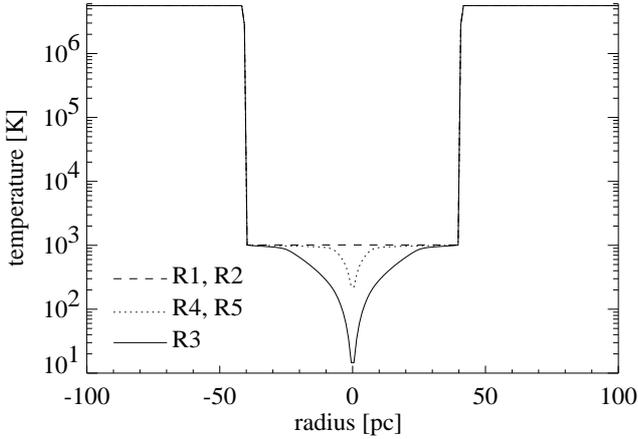,width=8.8cm,angle=0}
\caption{\label{fig04}Temperature profiles for the various
models. Models R1 and R2 show a homogeneous distribution in order to
fulfil pressure equilibrium with a homogeneous density
distribution. Pressure equilibrium is necessary also for model R3 with
a pronounced central density peak. The temperature profile for model R4
is calculated for hydrostatic equilibrium, model R5 has to fulfil
thermal equilibrium as well. This does not alter the temperature
distribution because of the chosen density profile.}
\end{figure}

$\Phi(r)$ is the gravitational potential.
The density and temperature profile of the cloud
is then calculated by integrating equations (\ref{eq304}) 
and (\ref{eq305}) from inside-out
using the core temperature of the cloud as a boundary condition and
truncating the cloud's outermost border where the energy density
of the cloud reaches the value of the hot plasma $e_{\rm{f}}$. This
condition is fulfilled for a radius of $R_{\rm{cld}} = 41$ pc.
The density distribution we get in this way
shows a clear core-halo structure with a tenuous rim and a
pronounced central core (Figs.~\ref{fig03} and \ref{fig04}) 
similar to observed interstellar clouds.

This cloud radius together with the hot plasma parameters leads 
to a global saturation parameter of $\sigma_0 = 4.88$ ($\Phi_s = 1$)
for which saturation of the heat flux is expected so that the 
classical mass-loss rate should be reduced for saturation effects 
according to CM77 by 51\% to a value of 
$5.6 \times 10^{-4}$ M$_{\odot}$ yr$^{-1}$ 
or according to DB93 by 67\% to a value of 
$3.7 \times 10^{-4}$ M$_{\odot}$ yr$^{-1}$, respectively.

In order to be close to the analytical considerations also a cloud
with homogeneous density distribution is treated, denominated 
as R1. With the same surface density and temperature as the stratified clouds 
it fulfills the condition of pressure equilibrium
with the ambient medium. The heat conduction is here implied only 
classically. Self-gravity, heating or cooling are neglected. 
For models R2 the density and temperature structures are not altered 
but the physics are extended in a way, that now the
effective heat flux is taken into account. 
These models can be used for a comparison with the analytical 
predictions of CM77 or DB93. 
For comparison reasons of all model clouds (also
of the homogeneous clouds) we fixed the radius to that of the 
R5 model, namely, 41 pc (Figs.~\ref{fig03} and \ref{fig04}) 
and, by this, the value of $\sigma_0$.

In a more advanced model (R3) the density distribution 
is then modified according to a core-halo structure, 
while the temperature distribution has also to be changed 
in order to guarantee pressure equilibrium (Fig.~\ref{fig04}). 
The mathematical form of the heat conduction process remains the same. 
As the further improvement for model R4 we apply the same 
density structure as in model R3 but take now self-gravity 
into account. The temperature profile has to change accordingly in
order to serve for hydrostatic equilibrium. And finally, as mentioned 
at the beginning of this part, R5 is developed as the most advanced
model implying also thermal equilibrium.

\section{Models}

Here we present the evolution of the five generally different models
and their variations with respect to the numerical resolution.

\subsection{Model R1a-R1b}

\begin{figure}[h]
\psfig{figure=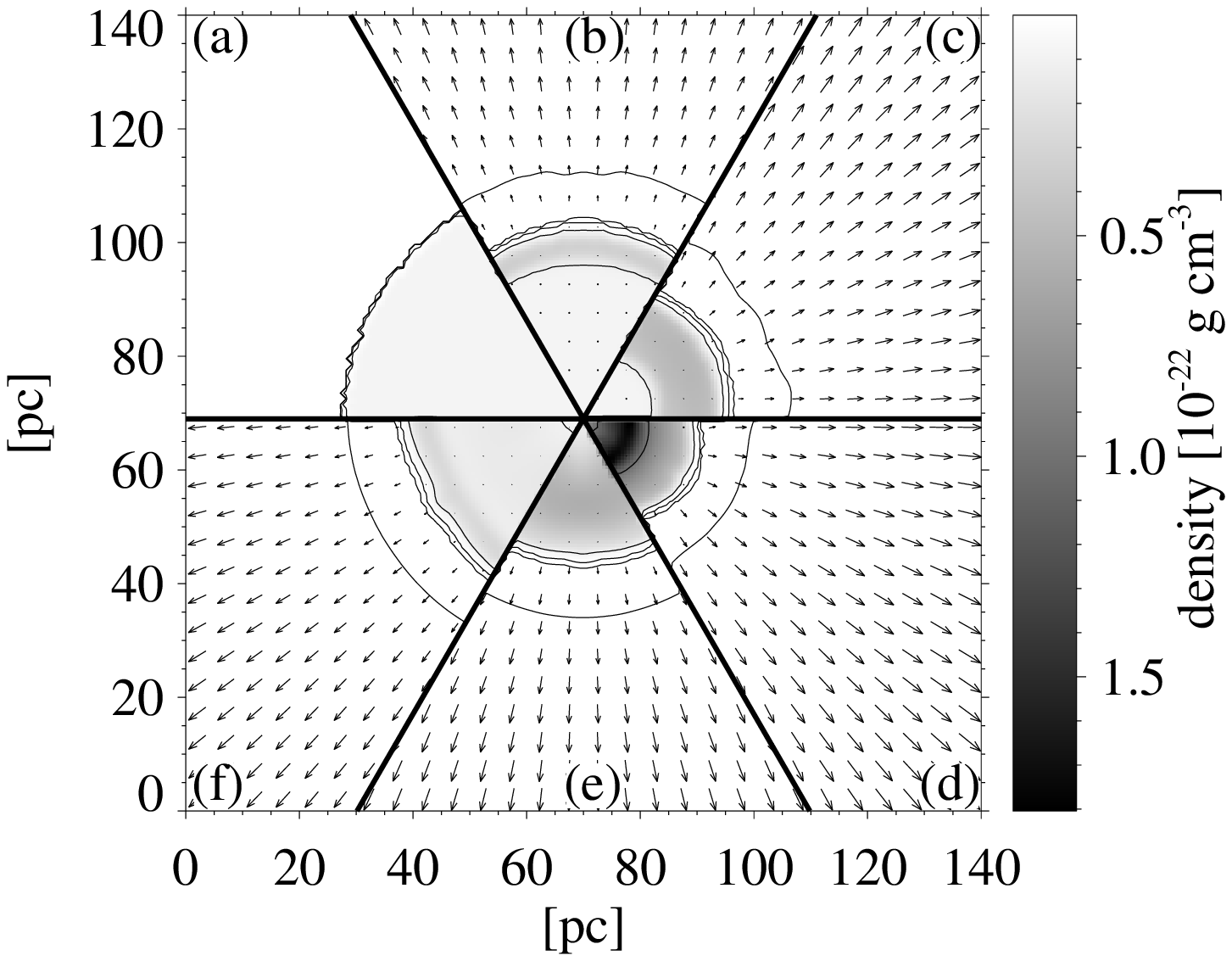,width=8.8cm}
\caption{\label{fig05}Evolution of the density distribution
of model R1a after 0 Myr (a), 1 Myr (b), 2 Myr (c), 3 Myr (d), 4
Myr (e) and 5 Myr (f). Lines of equal density are drawn between
$\rho~=~10^{-22}$ and $10^{-26}$~g~cm$^{-3}$ in steps of 1dex.} 
\end{figure}

\begin{figure}[h]
\psfig{figure=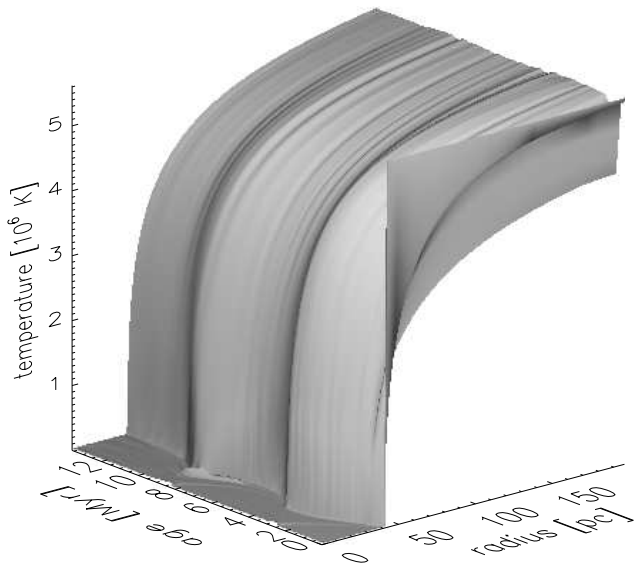,width=8.8cm,angle=0}
\caption{\label{fig06}Evolution of the temperature profile of model R1. 
The course is in good agreement with an analytically derived distribution for
$\sigma_0 < 1$ that takes only the classical heat flux into account.} 
\end{figure}

Although the global saturation parameter $\sigma_0$ suggests that saturation
effects in the heat flux are to be expected, this first simulation is performed
by only using the classical form of the heat flux. This is done in
order to investigate the consequence of this simple approach on the
mass-loss rate. The model starts with a homogeneous cloud of
$\rho = 8 \times 10^{-24}$ g cm$^{-3}$ and a spatially constant
temperature of $T = 1000$ K. The temporal 
evolution of the density distribution is shown in Fig.~\ref{fig05}.
The velocity field visualized by arrows in regions where $\rho <
10^{-25}$ g cm$^{-3}$ is directed away from the cloud during the whole
simulation. This is a clear indication for mass loss, while the inner parts of
the cloud begin to collapse. The trigger for this process is the huge heat
input at the cloud edge due to heat conduction, that rises the
temperature there and leads to an overpressure with a two-fold
consequence: First, it pushes material away
from the surface into the ICM whereby
the evaporated material is by far not as hot as the ICM when
it decouples from the cloud. So the resulting temperature
distribution outside the cloud is a rising function with
the cloud distance (see Fig.~\ref{fig06}).
Secondly, the evaporated
material generates a repulsion that leads to a preassure wave 
running through the cloud and further to a pulsation of the cloud as a
whole. The 
pulsation timescale corresponds very well to the dynamical
timescale of the cloud which is the sound crossing time. One has to
keep in mind that the oscillations are excited artificially by the
switch-on of the numerical simulation.

After some $10^5$ years a quasistatic temperature distribution has
formed (see Fig.~\ref{fig06}). In this figure also the pulsation of
the cloud can be revealed. A comparison of $T(r)$ at the end
of the calculation with an analytically derived distribution for
$\sigma_0 < 1$ (see Fig.~\ref{fig02}) shows a qualitatively good agreement.
The steep rise in temperature at the cloud surface and a smooth
asymptotic behaviour for large $r$ is typical for calculations with
only the classical heat flux.

The mass-loss rate is calculated far away from the cloud at a distance
of 95 pc using

\begin{equation}{\label{eq401}}
\dot{m} = 4 \pi r^2 \rho(r) v(r) \quad .  
\end{equation}

\begin{figure}[h]
\psfig{figure=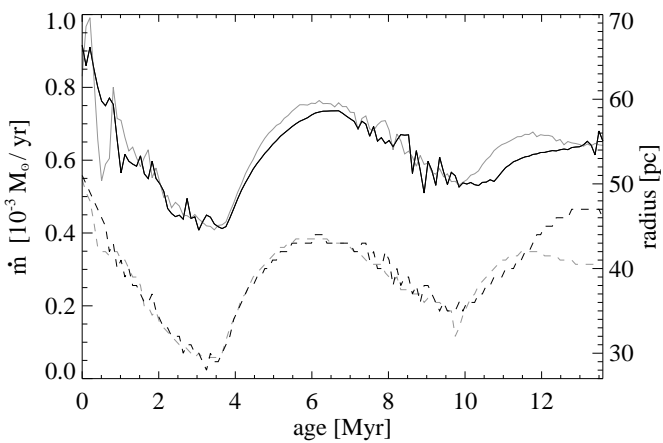,width=8.8cm,angle=0}
\caption{\label{fig07}Mass-loss rate for the models 1a (black solid
curve) and 1b  (grey solid curve) compared
with their cloud radius (lower black/grey dashed lines).} 
\end{figure}

Positive values of $\dot{m}$ stand for mass loss or evaporation of the
cloud while negative values of $\dot{m}$ indicate condensation of hot
ISM onto the cloud. During the whole simulation evaporation occurs
(see Fig.~\ref{fig07}) and its strength is well correlated
with the radius of the cloud as expected from eq.~\ref{eq207}.
$\dot{m}$ according to CM77 or DB93 requires a value of about 
$1 \times 10^{-3}$ M$_{\odot}$ yr$^{-1}$. This is in relatively good
agreement with the value derived from the simulations
 ($\dot{m} \approx$ 0.8$\times 10^{-3}$ M$_{\odot}$ yr$^{-1}$; see
table\ref{t2}). 
At the beginning $\dot{m}$ represents the analytical value, and this more
precisely for the spatially higher resolved simulation, what
demonstrates the accuracy of our treatment.
The model with higher resolution shows no significant
differences in its structure or evolution. Also the mass-loss rate is
compareable. This is an indication that we can handle heat condution
also in the low resolution case so that no time consuming highly resolved
simulations are necessary as long as one is not interested in surface
structural effects.

The limitation of heat conduction to the classical case leads to a very
large mass-loss rate in analytical studies as well as in numerical
simulations. Due to the enormous energy input the evaporated material
is accelerated to velocities near sonic speed.

\subsection{Model R2a-R2c}

\begin{figure}[h]
\psfig{figure=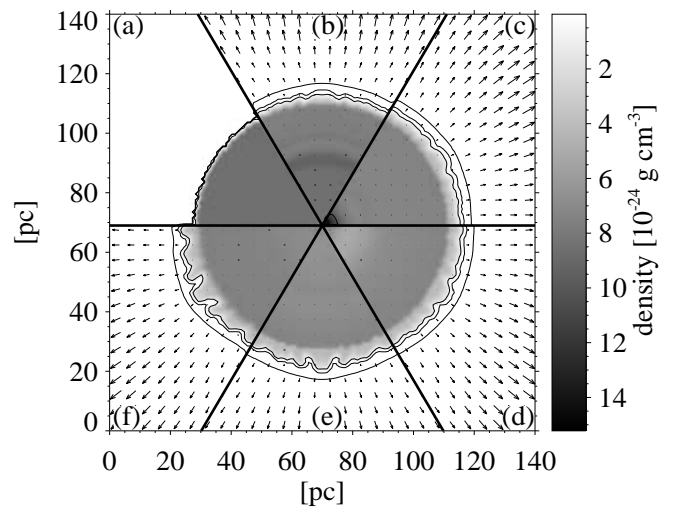,width=8.8cm}
\caption{\label{fig09}Evolution of the density distribution
of model R2a after 0 Myr (a), 5 Myr (b), 10 Myr (c), 15 Myr (d), 20
Myr (e) and 25 Myr (f). Lines of equal density are drawn between
$\rho~=~10^{-22}$ and $10^{-26}$~g~cm$^{-3}$ in steps of 1 dex.} 
\end{figure}

These models differ from R1 in that saturation effects are now taken
into account. The density and temperature
structure of the initial model is the same as in R1. 
The evolution in time of the density distribution is shown in
Fig.~\ref{fig09}.
Because of the limitation of the energy transport due 
to the saturated heat flux, also the energy input and with this the rise
in temperature at the cloud surface is reduced. The resulting
overpressure at the cloud rim is lower than in model R1. Nevertheless, a
pressure wave that runs through the cloud and causes density
fluctuations inside the cloud is triggered by the evaporated
material (see innermost denser shell in Fig.~\ref{fig09}b). 
In contrast to model R1, here no large-scale oscillations occur.

Also in this model cloud material is evaporated from the surface as 
a consequence of the additional heat input at the cloud surface due to heat
conduction. Unlike model R1 the overpressure is not large enough to
blow this material far and fast into space but it is mixed with the 
hot ICM and forms a transition layer around the cloud. 
The extension of this layer is visible
in Fig.~\ref{fig09} by the growth of the distances between iso-density
lines at the edge of the cloud. As time procedes Rayleigh-Taylor
(RT) instability grows in the transition layer because warm and dense
material is continuously accelerated into the hot rarefied plasma. 
The linear growth timescale for RT instability is 
$ \tau_{RT} = \sqrt{ \frac{ \lambda } {a} }$ 
where $\lambda$ is the size of the initial perturbation
which lies in the range of the cell size and $a$ is the effective acceleration. 
Assuming that the evaporated cloud material is accelerated to a value of 
0.05 Mach over a distance of 3 pc in the transition layer, 
$\tau_{RT}$ amounts to 0.8 Myrs which is in good agreement with the 
value derived from the simulation. 
RT instability would be also expected to arise in models R1, because
the conditions for instability are analytically fulfilled. However, 
the strong pulsations in this model are responsible to prevent its 
growth: phases in which RT instability is forced, i.e. when the cloud 
surface is accelerated outwards, are too short to allow its 
discernible growth.

\begin{figure}[h]
\psfig{figure=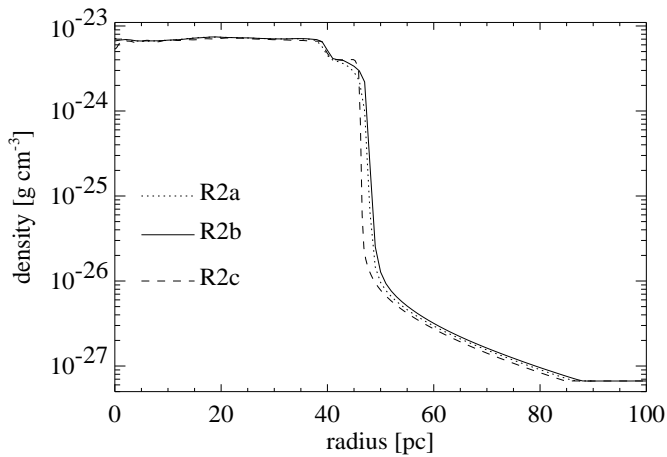,width=8.8cm,angle=0}
\caption{\label{fig10}Density profiles of the models R2a-R2c at a time 
$t = $ 30 Myr. The transition region for 40 pc $< r <$  48 pc is
clearly visible.} 
\end{figure}

The extension
of the transition layer is better visualized in Fig.~\ref{fig10} where
the density profiles of the different models R2a-R2c are plotted as a
function of the radius at time $t = $ 30 Myr. 
The transition layer is clearly visible between
40 pc $< r <$  48 pc. The outer edge of the layer cannot be sharply
defined. Also for larger radii up to 85 pc significant amounts of cloud 
material can be found. 

\begin{figure}[h]
\psfig{figure=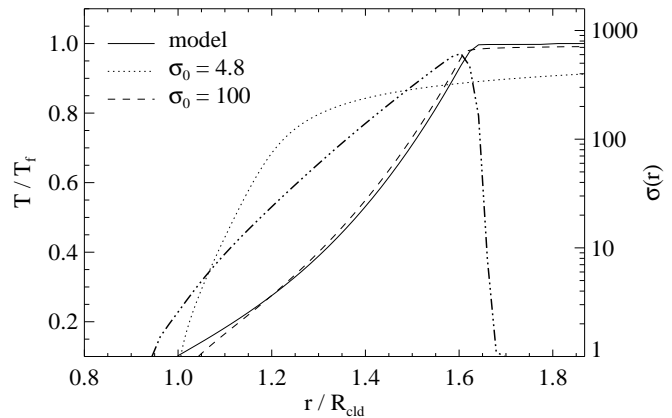,width=8.8cm}
\caption{\label{fig11}Comparison of different temperature profiles
between the model R2a after 62 Myr (solid curve) and theoretical
profiles of constant $\sigma_0$: 4.8 (dotted), $\sigma_0 =$ 100
(dashed). Additionally the local saturation parameter $\sigma(r)$ deduced 
from the simulation is plotted (dashed-dotted). 
The temperature is normalized to T$_f$ = 5.5 $\times 10^6$ K.}
\end{figure}

The results of the different R2 models show
only weak dependencies on resolution or boundary conditions.
In particular, the plateau between 40 pc and 45 pc in all R2 models in
Fig.~\ref{fig10} remains the same and reflects the regime 
where RT instability is excited.
The fluctuations inside the cloud result from the
pressure waves triggered by the repulsion of the evaporated
material. The temperature inside the transition zone is rising with
radius and has values of $T \approx 1.5 \times 10^3$ K at the inner 
boundary ($r =$ 40 pc) and reaches $T \approx 10^5$ K at $r =$ 48 pc. 
Because the density structure of the layer runs oppositely to the
temperature structure, the material in the transition zone can exist
almost in pressure equilibrium. Nevertheless, the remaining
slight pressure gradient at the outer boundary leads to the observed
evaporation. Like in the previous
model the velocity field is directed outwards and indicates
evaporation. In contrast to model R1, the maximum speed of the lost
material is subsonic with only 0.05 Mach ($\approx 21$ km s$^{-1}$). 
This mass-loss rate amounts to $\dot{m} \approx 1.4 \times
10^{-5}$ M$_{\odot}$ yr$^{-1}$ (see Fig.~\ref{fig14}) and is therefore
only 1/40 of the analytically predicted one and agrees at best for a
model with $\sigma_0 \approx 100$. 
We can show that the low Mach number is caused by the energy loss 
of the expanding (and evaporating) shell due to $p \cdot dV$ work against 
the external pressure.
From our R2 model one can derive that the specific thermal energy is reduced
by a factor of almost 500. 
This value can be derived from the density and temperature values
at the cloud edge (at 46 pc) and at $r=67$ pc where $T_f=5.6 \cdot 10^6$ K 
is reached. The density contrast amounts to $3 \cdot 10^{3}$ 
(see Fig.~\ref{fig10}), the temperature contrast to
$0.16$ (see Fig.~\ref{fig11}).  
Since the analytical consideration by CM77 cannot
account for this physical process, the evaporation in their model remains
at sound speed.
The slight differences in the simulations with different spatial 
resolutions and grid extension are not significant so that simulations with a
resolution of 1 pc cell$^{-1}$ are reliable. 
A comparison of the temperature profile of the simulation (see
Fig.~\ref{fig11}) with
an analytical one for $\sigma_0 = 4.8$ yields no good result whereas a
theoretical profile for $\sigma_0 = 100$ fits the simulated model
very well again.
Conclusively, the temperature profile is the most important factor for
the value of the evaporation rate and, on the
other hand, results from the local heat flux that is a function of
the density in the case of saturated heat conduction. With the
formation of the transition zone at the edge of the cloud the density
distribution has changed from the initial model and so does the
evaporation rate.

\subsection{Model R3}   

High-resolution observations of interstellar clouds, e.g. HVCs,
show that most of them consist of an inhomogeneous gas and dust distribution
with dense cores and rarefied envelopes. In order to analyze the
influence of the heat conduction process on more realistic core-halo
structures, the following cloud model consists of a pronounced
core without self-gravity or heating and cooling. The temperature
distribution of the initial model was chosen to account for pressure
equilibrium. In the core region with some 10 particles cm$^{-3}$ the
temperature drops to about 10 K. Towards the edge it rises to $T =
300$ K at $r = 10$ pc and $T = 1000$ K at $r = 30$ pc. Because the
heat flux is very small in the inner regions as a consequence of the
low temperature, its influence will be limited to the cloud edge. 
\begin{figure}[h]
\psfig{figure=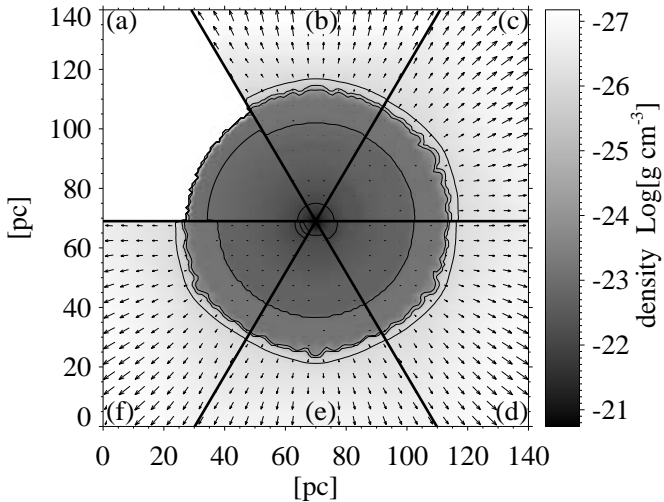,width=8.8cm}
\caption{\label{fig12}Evolution of the density distribution
of model R3 after 0 Myr (a), 5 Myr (b), 10 Myr (c), 15 Myr (d), 20
Myr (e) and 25 Myr (f). Lines of equal density are drawn between
$\rho~=~10^{-22}$ and $10^{-26}$~g~cm$^{-3}$ in steps of 1dex.} 
\end{figure}
The evolution of the density distribution is shown in Fig.~\ref{fig12}.
Similar to model R2, a transtion region forms that is visible by the
growth of the separation of density isolines at the edge of the
cloud. Also the growth of RT instabilies is discernible. 

\begin{figure}[h]
\psfig{figure=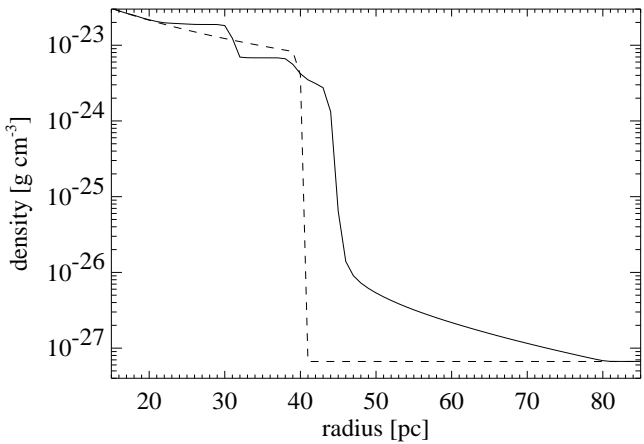,width=8.8cm,angle=0}
\caption{\label{fig13}Density profile of the model R3 at a time 
$t = $ 37 Myr (solid curve) in comparison with the initial model (dashed).} 
\end{figure}

\begin{figure}[h]
\psfig{figure=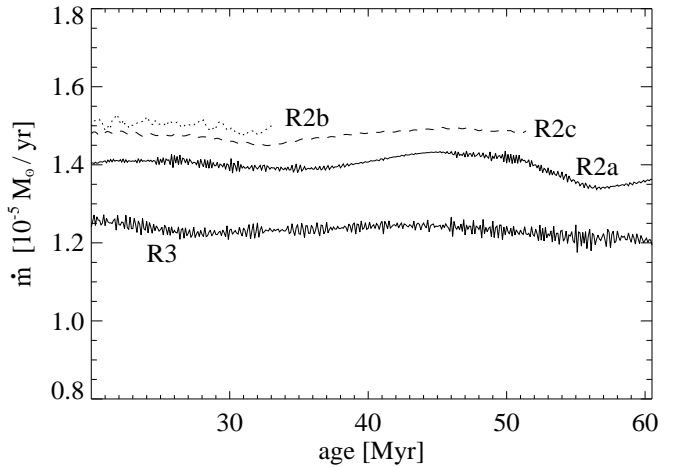,width=8.8cm,angle=0}
\caption{\label{fig14}Evaportion rates for the models R2a-R2c and R3.} 
\end{figure}

In order to analyze the evolution at the cloud rim, a density profile
at $t = 37$ Myr is plotted in Fig.~\ref{fig13} and compared with the
initial model.
Clearly visible is the transition region with similar expansion like
model R2. 
This expansion excites sound waves travelling inwards, by this, 
dissipating energy, and rising the temperature by about 150 K in the 
region 30 pc $< r <$ 40 pc whereas the core region remains nearly unaffected.
%The pressure wave triggered by the repulsion of the evaporated material 
%affects the core region only marginally. Due to the induced
%overpressure, cloud material is transported towards the cloud core
%until a new pressure equilibrium is reached. The density
%distribution is now quasistatic. The cloud can be devided into three
%parts: The innermost central part with $r < 30$ pc is characterized by
%its huge density and low temperature. The second part with 30 pc $< r
%<$ 40 pc is separated from the inner part by a density jump
%and is characterized by a nearly homogeneous density and temperature
%distribution. The temperature in this region is still high enough so
%that heat conduction can smear out density inhomogeneities. The third
%and outermost part of the cloud is formed by the transition layer that
%connects the cloud with the ICM. 
The cloud's density distribution becomes quasistatic after the
redistribution of the
density so that a new pressure equilibrium is reached. The cloud structure 
is now comparable to model R2 except the fact, that the innermost part 
with $r < 30$ pc is almost inert and characterized by its huge density and
low temperature. Also 
the density distribution inside the transition layer is
comparable to the one of model R2 (see Fig.~\ref{fig10}
and~\ref{fig13} for comparison). Because the evaporation rate is
strongly determined by the processes in this transition layer, it is
expected, that the mass-loss rate of both models should be the
same (see Fig.~\ref{fig14}). 
%Fig.~\ref{fig12} indicates also, that no
%large-scale fluctuations of the density inside the cloud occur.
%
%
%The mass-loss rate of model R3 shows no large-scale fluctuations
%because a quasi-stationary density distribution is installed. In model R2,
%density fluctuations inside the homogeneous cloud are visible during
%the whole simulation. 
The reason for the slightly lower 
evaporation rate for model R3 can be found by the fact, that evaporated
material can be provided only from the homogeneous part of the cloud
within 30 pc $< r <$ 40 pc while the innermost part of the cloud is inert
against heat conduction. Conclusively no significant differences
can be found in the evolution between a multi-phase cloud and a 
homogeneous cloud.

\subsection{Model R4}

Clouds like those simulated in model R3 are massive enough so that
self-gravity is no longer negligible. Model R4 was therefore
calculated with self-gravity by solving the Poisson
equation for the corresponding density distribution. The temperature
of the initial model was set for hydrostatic
equilibrium. The dense core has therefore to achieve a higher temperature 
than model R3 in order to prevent the cloud from gravitational
collapse. The core temperature is about $T = 230$ K, rises up to
$T = 920$ K at $r = 10$ pc and reaches $T = 1000$ K at $r = 30$ pc. 
Heat conduction is very sensitive to a rise in temperature ($\kappa
\propto T^{5/2}$, see eq.~\ref{eq202}) so that energy transport
into the core region is expected.

\begin{figure}[h]
\psfig{figure=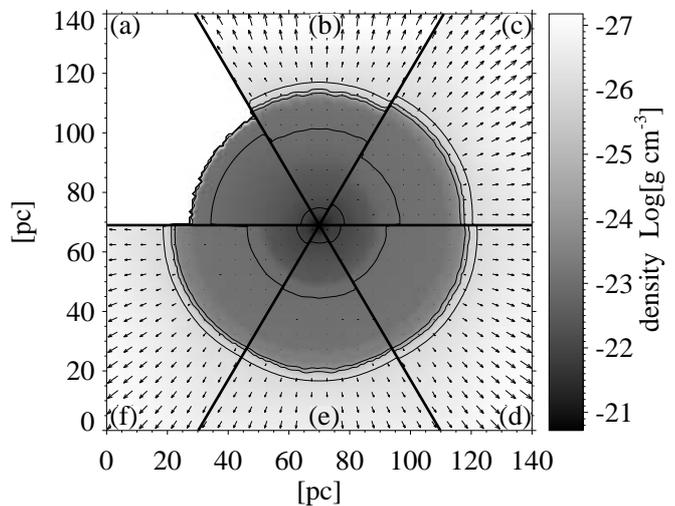,width=8.8cm}
\caption{\label{fig15}Evolution of the density distribution
of model R4 after 0 Myr (a), 5 Myr (b), 10 Myr (c), 15 Myr (d), 20
Myr (e) and 25 Myr (f). Lines of equal density are drawn between
$\rho~=~10^{-22}$ and $10^{-26}$~g~cm$^{-3}$ in steps of 1dex.} 
\end{figure}

The evolution of the density distribution is shown in Fig.~\ref{fig15}.
A transition region at the cloud edge has formed with an extension
almost constant after a formation time of about 15 Myr. The
structure of this layer is much more regular than in the previous
models without self-gravity where RT instabilities
dominate the region. Self-gravity is a process that stabilizes the
cloud against RT instability (Murray et al. \cite{mu93}): 
the effective acceleration which the surface material experiences
as the net effect of evaporation (like in models R2 and R3) vs.
self-gravity points radially inwards as the density gradient does.

\begin{figure}[h]
\psfig{figure=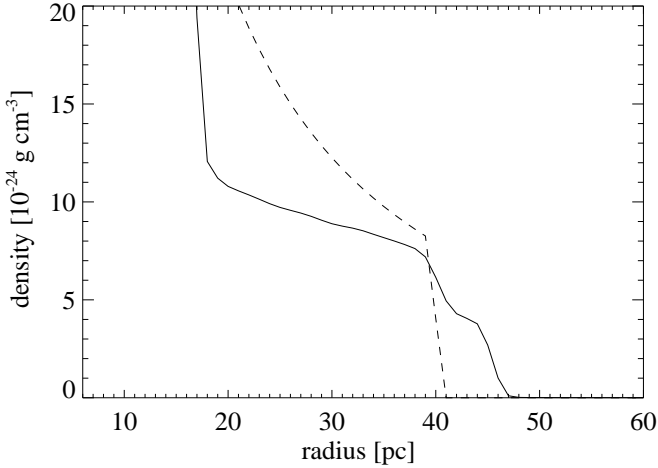,width=8.8cm,angle=0}
\caption{\label{fig16}Density profile of the model R4 at a time 
$t = $ 53 Myr (solid curve) in comparison with the initial model (dashed).} 
\end{figure}

\begin{figure}[h]
\psfig{figure=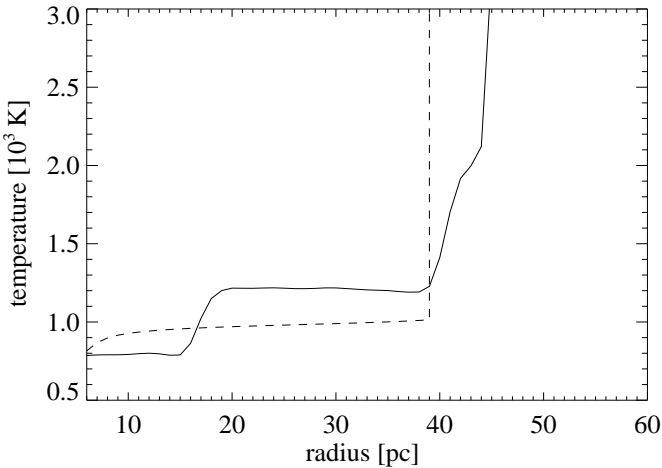,width=8.8cm,angle=0}
\caption{\label{fig17}Temperature profile of the model R4 at a time 
$t = $ 53 Myr (solid curve) in comparison with the initial model (dashed).} 
\end{figure}

Similar to model R3 the cloud can be devided into an inert core, a
  heat-conduction dominated zone with nearly homogeneous density 
distribution and the transition zone.
They are
visible by the density profile after 53 Myr in comparison
to the initial model (Fig.~\ref{fig16}) and the corresponding
temperature profiles (Fig.~\ref{fig17}). 
%
%For cloud regions with $r <$ 18 pc the situation is the same as in the 
%innermost region of model R3. This area is nearly inert against 
%heat conduction and is separated
%from the heat-conduction dominated region within 18 pc $< r <$ 40 pc by a
%density and temperature discontinuity. Heat conduction tends to
%homogenize the temperature distribution. 
Because of gravitation the density
profile of the intermediate zone is not as flat as the temperature profile 
but approaches hydrostatic equilibrium. 
%The third zone, the transition layer, is
%similar to the previous models.

\begin{figure}[h]
\psfig{figure=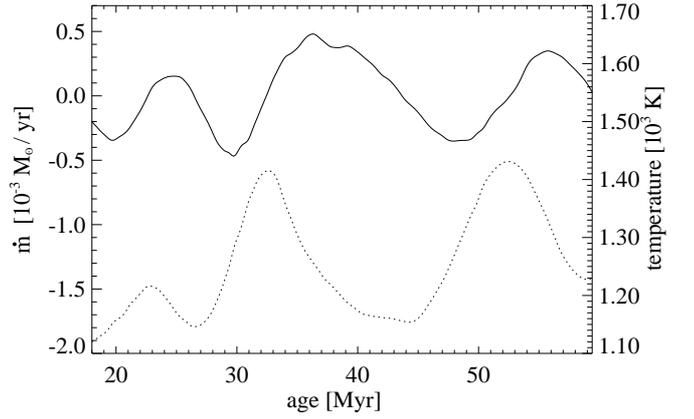,width=8.8cm}
\caption{\label{fig18}Evolution of the mass-loss rate (solid curve)
and the temperature (dotted) at the inner boundary of the transition
zone ($r =$ 40 pc) for model R4.} 
\end{figure}

\begin{figure}[h]
\psfig{figure=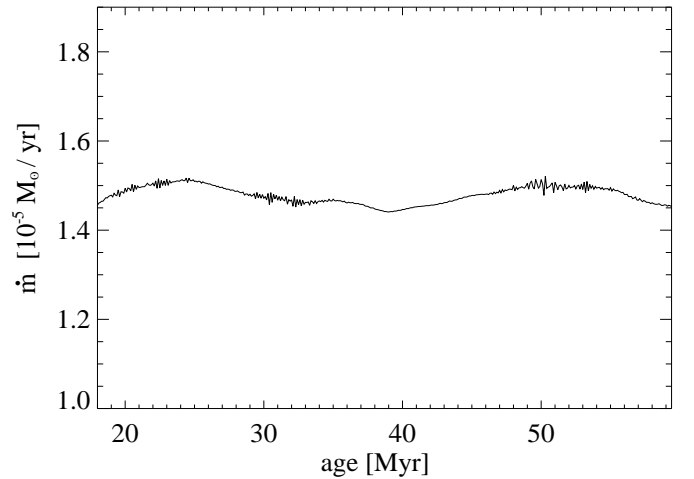,width=10.0cm,angle=0}
\caption{\label{fig19}Evaportion rate for the model R4.} 
\end{figure}

During the whole simulation energy is transported into the cloud at
$r >$ 18 pc due to heat conduction. This energy flow is connected with a
mass transport because of the induced overpressure. This mass flow is
neither locally nor temporally constant but corresponds to the
temperature of the transition zone. To visualize this correlation, the
mass flow and the temperature at $r =$ 40 pc (inner boundary of
the transition zone) are plotted in Fig.~\ref{fig18}. 
Similar to the previous models the evaporation of cloud material leads
to a repulsion and therefore triggers a pressure wave inside the
cloud. The consequence
of the inward-travelling wave, namely, the negative $\dot{m}$  
is visible in Fig.~\ref{fig18} (e.g.: $t =$ 27 Myr) and correlated
with a rise in temperature of the
transition zone. The wave interacts with the density discontinuity
at $r = 18$ pc (e.g.: $t =$ 30 Myr), is reflected, travels outwards,
reenters the transition zone and, with that,
pushes cold cloud material into the transition zone (e.g.: $t =$ 32.5 Myr). 
While material continuously flows into the transition layer the
temperature there is further reduced (e.g.: 32.5 $< t <$ 42 Myr).
This transport is stopped as soon as a new hydrostatic equilibrium
has been adjusted. The temperature fluctuations at the inner boundary
of the transition zone also become visible at the temperature
distribution outside the cloud by means of the evaporated mass
flow. E.g. at $r =$ 95 pc the temperature
varies by 2\% correlated with the evaporation rate of about $1.5 \times
10^{-4}$ M$_{\odot}$ yr$^{-1}$ on average (see Fig.~\ref{fig19}). 
This is slightly larger than in model
R3 because the area feeding the evaporation is
larger than in the previous model. 

This study shows that dynamical processes inside the cloud must not be
neglegted but determine the transport inside the transition zone
and with this the temperature profile at the cloud edge. Even far away
from the cloud this dynamics can be comprehended by the mass-loss rate
or temperature distribution. Because of the much larger sound velocity
than that of the evaporation flow the external temperature fluctuation
affect the evaporaton rate.

\subsection{Model R5}

Because already 
the studies by McKee \& Cowie (1977) indicate that a
reduction of the heat flux due to cooling processes is expected but only
for clouds with $\sigma_0 < 0.027$, model R5 takes
heating and cooling processes into account. 
Opposite to CM77 this model is calculated with self-gravity.  
From model R4 we conclude that self-gravity acts only as a smoothing 
agent for reducing instabilities that occur when heated material is  
evaporated from the cloud surface into the ICM. The mass-loss rate  
was not significantly altered by this process. So, this model should be  
comparable to the analytic model of Cowie \& McKee.
In models R2-R4 heat
conduction leads to a temperature and therefore to a pressure increase
at the edge of the cloud where the transition zone forms between the
cold cloud and the hot ICM. In order to compensate the local energy
input, two processes are physically plausible: The cloud distributes
the energy throughout its volume by dynamical transport processes
and/or it emits the energy by the process of radiative cooling. 
Cooling is very sensitive to the particle density and is
most effective in regions with high density. The question is
therefore whether the cooling efficiency is large enough to alter the
formation or evolution of the transition zone and, with that, to change
the evaporation rate.

\begin{figure}[h]
\psfig{figure=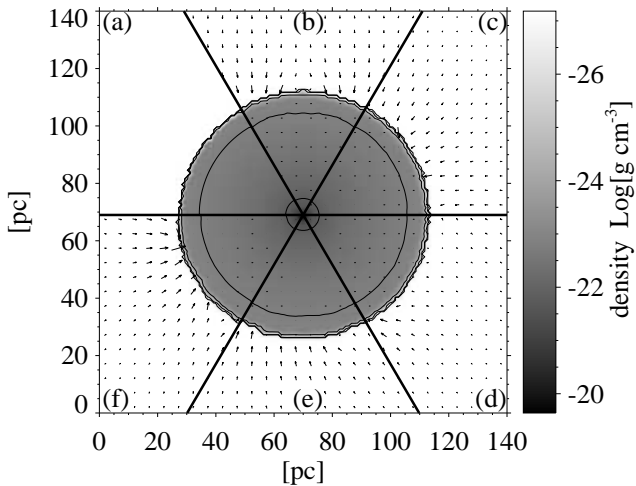,width=10.0cm}
\caption{\label{fig20}Evolution of the density distribution
of model R5 after 0 Myr (a), 5 Myr (b), 10 Myr (c), 15 Myr (d), 20
Myr (e) and 25 Myr (f). Lines of equal density are drawn between
$\rho~=~10^{-22}$ and $10^{-26}$~g~cm$^{-3}$ in steps on 1dex.} 
\end{figure}

There are two main differences visible in the evolution of the density
distribution of model R5 (Fig.~\ref{fig20}) in comparison to the
previous models.
First of all, the velocity field outside the cloud
is directed towards the cloud that clearly indicates accretion of ICM
onto the cloud's surface. Secondly, the density distribution inside
the cloud remains nearly unaffected by heat conduction. Similar to the
models R2-R4 a transition zone is formed but its extension is 
much narrower. This means that the influence of heat conduction is
restricted to only a thin layer around the cloud with a thickness of only 
few parsecs.

\begin{figure}[h]
\psfig{figure=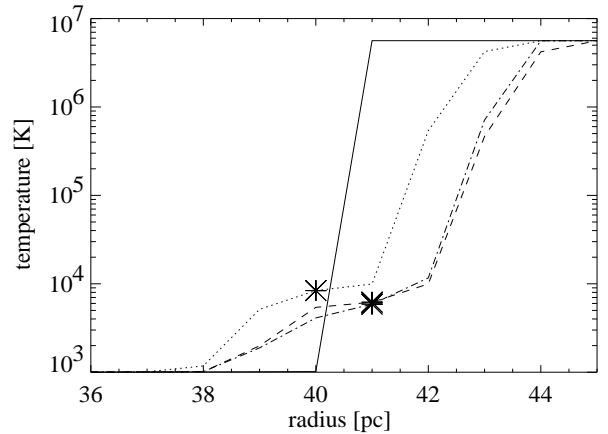,width=8.8cm,angle=0}
\caption{\label{fig21}Comparison of temperature profiles of model R5
at different ages: initial model (solid curve), 
after 30 Myr (dotted), after 40 Myr (dashed) and after 55 Myr (dashed-dotted). 
The radii ($r_{\star}$) at which the energy input due to heat conduction 
plus radiative heating is balanced by radiative cooling are marked by crosses.} 
\end{figure}
The used cooling function shows a steep increase 
for $T$ around $10^4$ K. For regions with temperatures below
$10^4$ K this means that they can be heated up to about $10^4$
K with high efficiency but then should somewhere easily find an
energetic balance between heating and cooling.
This equilibrium situation is determined by the Field length 
$\lambda_{\mbox{\tiny F}}$ (Eq.~\ref{eq208}). In the outermost shell 
of the cloud $\lambda_{\mbox{\tiny F}}$ amounts to 
$1.4 \times 10^3 \, R_{\rm{cld}}$ 
(Vieser \& Hensler 2005; hereafter: Paper II). For this 
reason the occurence of RT instability is prevented and no further refined
grid resolution is neccesary. Due to the density gradient towards the cloud
core $\lambda_{\mbox{\tiny F}}$ drops within an interface formed by 
heat conduction down to values much smaller than $R_{\rm{cld}}$. 
Because of its high cooling efficiency by larger density a temperature 
of 8000-10$^4$ K is achieved at which its thermal energy
density is not exceeding the gravitational one.
As a consequence, the interface does not dissolve from the
cloud but accumulates surrounding gas that enters the interface by means
of heat conduction.

From Fig.~\ref{fig21} one can recognize that between 30 Myrs and 40
Myrs a quasi-stationary temperature distribution has been adjusted. 
The balance at which the energy input due to heat conduction plus heating 
is compensated by cooling is reached at temperatures of $10^4$ K and radii 
$r_{\star}$. 
For the reached values at $r_{\star}$ the Field length amounts to not
more than $10^{-2} \, R_{\rm{cld}}$. 
This effect has two consequences: At first, 
the physical one is that 
the parameter regime of condensation is reached; secondly, 
the cell size is larger than the typical thermodynamical length scale, 
i.e.\ here the mean free path of electrons and ions, respectively.
Therefore, this means that the hydrodynamical treatment is appropriate 
(\cite{ki04}).

\begin{figure}[h]
\psfig{figure=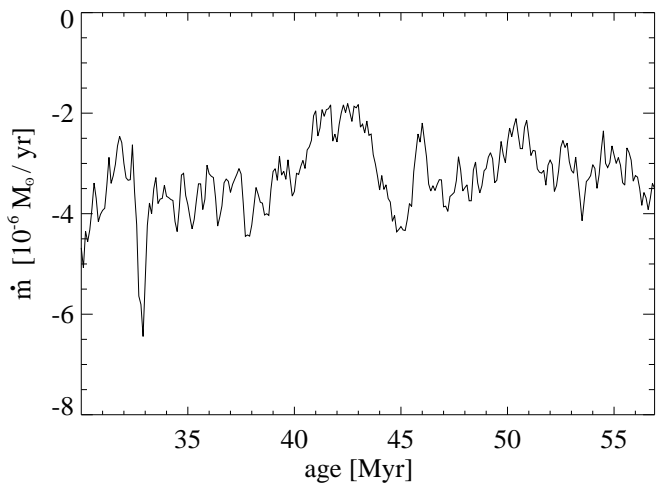,width=8.8cm,angle=0}
\caption{\label{fig22}Evolution of the mass-loss rate of model R5.} 
\end{figure}

Thermal energy is transported from  $r > r_{\star}$ inwards during the
whole calculation and finds an equilibrium with the cooling rate at
$r_{\star} \approx$ 41 pc and 8000 K. Consequently,
condensation occurs (Fig.~\ref{fig22}) in this model with a value of
about $-3.5 \times 10^{-6}$ M$_{\odot}$ yr$^{-1}$.
\begin{figure}[h]
\psfig{figure=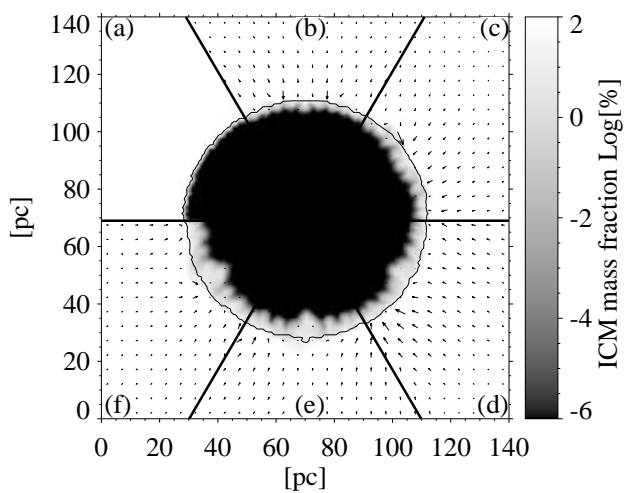,width=10.0cm}
\caption{\label{fig23}Evolution of the ICM mass fraction
($\rho_{\rm{ISM}} / \rho_{\rm{cld}}$)
of model R5 after 0 Myr (a), 10 Myr (b), 20 Myr (c), 30 Myr (d), 40
Myr (e) and 50 Myr (f). The solid line represents the cloud boundary.} 
\end{figure}
Although the available conductive energy at $r < r_{\star}$ is reduced
the temperature rises moderately above $10^3$ K for $r >$ 38 pc
(Fig.~\ref{fig21}), by this generating a turbulent velocity field that
transports the accreted material from the cloud surface towards deeper
layers. Due to the condensation of ICM onto the cloud and
its transport inside the cloud, the mass fraction of ICM is rising
continuously. The distribution of ICM in comparison to the whole
cloud mass is visualized in Fig.~\ref{fig23}.
A growing shell becomes visible with an ISM mass fraction of at least
1\%. After 50 Myr this layer has reached a thickness of about 10 pc. 

To emphasise, heat conduction together with cooling
processes does not only change the mass flow from evaporation to
condensation but also provide a new possibility to enrich the cloud
with ICM and to distribute it throughout large parts of the cloud.

\section{Discussion}

In this paper we investigate the evolution of interstellar clouds
that are resting in a hot rarefied medium. Models with different
density structures and various
physical processes are compared with the aid of numerical
simulations. The classical evaporation rate
without taking saturation effects into account amounts to 10$^{-3}$
M$_{\odot}$ yr$^{-1}$. Analytical consideration of saturation allows
to lower it to about 
$5.6 \times 10^{-4}$ M$_{\odot}$ yr$^{-1}$ (CM77) or 
$3.7 \times 10^{-4}$ M$_{\odot}$ yr$^{-1}$ (DB93), respectively. 

\begin{table}[h]
\centering
\caption{\label{t2}Mass-loss rates of the various simulations 
(negative values mean mass accretion!) }
\begin{tabular}{cc}
Model 
& $\dot{m}$ [M$_{\odot}$ yr$^{-1}$] \\ \hline 
R1 & $0.8 \times 10^{-3}$ \\
R2 & $1.4 \times 10^{-5}$ \\
R3 & $1.2 \times 10^{-5}$ \\
R4 & $1.4 \times 10^{-5}$ \\
R5 & $-3.5 \times 10^{-6}$ \\
\end{tabular}
\end{table}

The global saturation parameter of heat conduction in
all calculations here was $\sigma_0 = 4.88$.
The mass-loss rates
deduced from the simulations are summarized in Table~\ref{t2}.
In order to compare with analytical studies (CM77, DB93), the
initial model R1 shows a homogeneous density distribution and
a temperature determined by pressure equilibrium. 
Influences by self-gravity or heating and cooling are neglected. 
As a first step, in model R1 only the classical description of 
heat conduction is taken into account. 
{\it The mass-loss rate} is in good agreement with the analytical 
results, but we learn that it  {\it oscillates due to excited 
cloud pulsations}.

The evolution of a cloud taking saturation effects into account
(model R2) proceeds not as violent as in R1. The energy input
into the cloud is reduced due to the saturated heat flux. Nevertheless
the cloud's outermost layer is heated up and forms a
transition layer between the cold cloud and the hot ICM 
in which the material can remain near pressure equilibrium with the
ICM. From there cloud gas is dissolved and pushed into the hot
ICM at subsonic speed.
This, however, changes the temperature and density profiles in the 
cloud's environment drastically in comparison to the initial model, so
that it is not surprising that the mass-loss rate is reduced to about
one order of magnitude below the analytical value (CM77, DB93). On the
other hand, this value is similar to the one derived analytically
for $\sigma_0 \approx 100$ and indeed the mean local saturation
parameter deduced from the simulated temperature profile yields about
100 in contrast to $4.88$ set as the formal value. 
From that we learn, that {\it the global
saturation parameter is therefore not conclusive 
to derive the mass-loss rate.}

The influence of a multi-phase density structure of the cloud on the
evaporation rate is not significant (model R3). Since density and
temperature at the cloud surface are similar to the previous model
which serve as the dominant factors for the heat input and by this 
for the mass loss rate, the same result is expected.

That RT instability is the responsible mechanism that shapes the 
clouds' surfaces in models R2 and R3 becomes plausible, 
because in model R4 the acceleration by self-gravity overcomes
the outwards directed acceleration of the expanding shell. 
Thus, the shape of the cloud remains undisturbed but 
the mass-loss rate is slightly higher than in model R3 because 
the temperature near the cloud core of the initial
model is higher for hydrostatic equilibrium. By this,
heat conduction can affect deeper layers of the cloud.

In model R5 heating and cooling processes are added. Since the 
used cooling function increases steeply at $10^4$ K the heating 
due to heat conduction can be balanced by radiative cooling at 
around $10^4$ K for sufficiently high densities. In the presented 
cloud model saturated heat conduction is not able to transport
enough thermal energy into the cloud in order to rise the temperture
above $10^4$ K and therefore to evaporate the cloud. The cloud rim
remains at temperatures of about $10^4$ K. Correlated with the energy
transport onto the cloud's surface, mass is accumulated from the ISM
and condenses onto the cloud. The cloud regions near
the surface are slightly heated so that a turbulent velocity field 
emerges that mixes the accreted ICM into deeper layers. 

Three major conclusions can be drawn from the numerical models presented
in this paper: 
1) {\it Evaporation itself alters the environmental
conditions} in the sense of positive feedback. 
2) The assumption of {\it classical heat conduction is invalid under real 
conditions} of interstellar clouds. In contradiction to those analytical 
results, heat conduction is reduced to the saturated limit. 
3) In connection with {\it radiative cooling} this {\it can lead to 
condensation of hot ICM onto clouds} in a regime of cloud parameters 
where evaporation is requested from the analytical approach. By this it 
provides an alternative way to accrete and mix ICM with cloudy material.

\begin{acknowledgements}
The authors thank Tim Freyer for stimulating discussions and Tomek
Plewa for providing us with his numerical code solving the 
heat conduction equation. We gratefully acknowledge suggestions 
by an anonymous referee which helps for clarification of the results.

This work was partly supported by the
Deutsche Forschungsgemeinschaft (DFG) under grant numbers He~1487/5-3
and He~1487/25-1. The computations were performed at the Rechenzentrum
der Universit\"at Kiel, the Konrad-Zuse-Zentrum f\"ur Informationstechnik
in Berlin, and the John von Neumann-Institut f\"ur Computing in
J\"ulich.
\end{acknowledgements}

%__________________________________________________________________

\Online

\begin{appendix}
\section{Heat conduction - implementation \& tests}

The implementation of heat conduction into the existing and tested hydro-code
turned out to be more complicated than expected in order not to limit 
the hydro-timestep $\tau_{\rm{hyd}}$ to the conduction-timestep given by
\begin{equation}{\label{eqa01}}
\tau_{\rm{cond}} = \frac{e}{T \kappa} \left[ {\rm{min}}(dr , dz) \right]^2
\end{equation}
where $dr$ and $dz$ are the distances between the centers of two
neighbouring cells in $r$ and $z$ direction respectively. The problem
dealing with these incompatible timescales is treated by using a semi-implicit
scheme to solve the heat conduction equation. 

The equation describing the change of the internal energy density due
to heat conduction
\begin{equation}{\label{eqa02}}
\frac{\partial e}{\partial t} =  
\frac{1}{r} \, \frac{\partial}{\partial r}
\left[ r \kappa \frac{\partial T}{\partial r} \right] +
\frac{\partial}{\partial z}
\left[ \kappa \frac{\partial T}{\partial z} \right]
\end{equation}
is transformed via the equation of state for an ideal gas into an
expression for the change of the temperature:
\begin{equation}{\label{eqa03}}
\frac{\partial T}{\partial t} = \frac{T}{e}
\frac{1}{r} \, \frac{\partial}{\partial r}
\left[ r \kappa \frac{\partial T}{\partial r} \right] +
\frac{T}{e}
\frac{\partial}{\partial z}
\left[ \kappa \frac{\partial T}{\partial z} \right] \quad .
\end{equation}
Both spatial directions are treated seperately using the same strategy.
Therefore only the solution of the radial part is described in the
following: The new temperature distribution is calculated for
all cells in the $r$-direction with a fixed $z$-coordinate. This is
done using an ADI-method. After the temperature in this row has
converged, the index of the $z$-coordinate is increased by one and the equation
is solved for the next row. 

\begin{figure}[h]
\psfig{figure=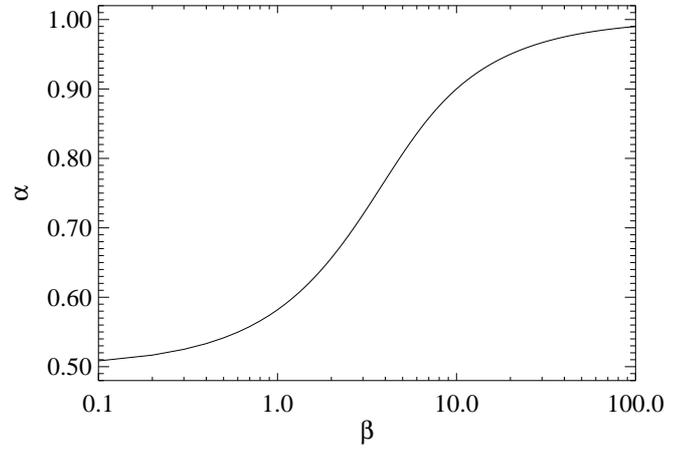,width=8.8cm}
\caption{\label{figa01}Weightning factor $\alpha$ as a function of $\beta$.}
\end{figure}

If the discretisation of eq.~\ref{eqa03} is 
done in a pure explicit way not only the timestep is limited to the
conduction-timestep. This procedure by itself is numerically unstable
(Crank~\&~Nicolson~\cite{cn47}). Instead of this approach we used
a semi-implicit procedure where the implicit part is weighted by
a factor $0.5 < \alpha < 1$ and the explicit part by a factor $1 - \alpha$
what results in:
\begin{eqnarray}{\label{eqa04}} 
\frac{ T_{\rm{i,j}}^{\,\rm{n+1}} -  T_{\rm{i,j}}^{\,\rm{n}} }
{\Delta t} 
& = & 
(1-\alpha) \; 
\left\{ \,
\frac{2}{r_{\rm{i,j}}+r_{\rm{i,j+1}}} \; 
\frac{ T_{\rm{i,j}}^{\,\rm{n}} } { e_{\rm{i,j}}^{\,\rm{n}} } \;
\frac{1}{\Delta r} \; \times \right. \nonumber\\
& &
\left[ 
\left( 
r_{\rm{i,j+1}} \; \kappa_{\rm{i,j+1}}^{\,\rm{n}} \; 
\frac{ T_{\rm{i,j+1}}^{\,\rm{n}} - T_{\rm{i,j}}^{\,\rm{n}} }
{\Delta r}
\right)
- \right.
\nonumber\\
& &
\left.
\left.
\left( 
r_{\rm{i,j}} \; \kappa_{\rm{i,j}}^{\,\rm{n}} \; 
\frac{ T_{\rm{i,j}}^{\,\rm{n}} - T_{\rm{i,j-1}}^{\,\rm{n}} }
{\Delta r}
\right)
\right]
\right\} \; + \nonumber\\
& & 
\alpha \; 
\left\{ \,
\frac{2}{r_{\rm{i,j}}+r_{\rm{i,j+1}}} \; 
\frac{ T_{\rm{i,j}}^{\,\rm{n+1}} } { e_{\rm{i,j}}^{\,\rm{n+1}} } \;
\frac{1}{\Delta r} \; \times \right. \\
& &
\left.
\left[ 
\left( 
r_{\rm{i,j+1}} \; \kappa_{\rm{i,j+1}}^{\,\rm{n+1}} \; 
\frac{ T_{\rm{i,j+1}}^{\,\rm{n+1}} - T_{\rm{i,j}}^{\,\rm{n+1}} }
{\Delta r}
\right)
- 
\right.
\right.
\nonumber\\
& &
\left.
\left.
\left( 
r_{\rm{i,j}} \; \kappa_{\rm{i,j}}^{\,\rm{n+1}} \; 
\frac{ T_{\rm{i,j}}^{\,\rm{n+1}} - T_{\rm{i,j-1}}^{\,\rm{n+1}} }
{\Delta r}
\right)
\right]
\right\} \quad . \nonumber 
\end{eqnarray}
This equation can be rearranged and the different coefficients be
combined to form: 
\begin{eqnarray}{\label{eqa05}}
A_{\rm{i,j}} \; T_{\rm{i,j-1}}^{\,\rm{n+1}} +
B_{\rm{i,j}}  \; T_{\rm{i,j}}^{\,\rm{n+1}} +
C_{\rm{i,j}} \; T_{\rm{i,j+1}}^{\,\rm{n+1}}
= D_{\rm{i,j}} \quad .
\end{eqnarray}
For a row with the index $i$ kept constant this describes a
maxtrix equation which can be solved with standard methods (e.g.
Press~et~al.~\cite{pe86}). We used an ADI-method. The weightning 
parameter $\alpha$ is chosen in a manner so that the implicit part is
pronounced if the conduction-timestep is much smaller than the
hydro-timestep. This means that high temperatures or large
conductivities dominate the integration domain. The ratio between the
two timesteps is defined as $\beta$
\begin{eqnarray}{\label{eqa06}}
\beta = \frac{ \tau_{\rm{hyd}} }{ \tau_{\rm{cond}} } \quad ,
\end{eqnarray}
that deals as a quantity to derive $\alpha$:
\begin{eqnarray}{\label{eqa07}}
\alpha = \frac{ 1 - \beta - \exp \left[ - \beta \, \right] }
{ \beta \left( \exp \left[ - \beta \, \right] - 1 \right) } \quad ,
\end{eqnarray}
with the additional limitation
\begin{eqnarray}{\label{eqa08}}
\alpha = \left\{
\begin{array}{r@{\quad:\quad}c}
0.5 & \alpha < 0.5 \\
\alpha & 0.5 \le \alpha \le 1 \\
1   & \alpha > 1
\end{array}
\right.
\end{eqnarray}
so that the minimal implicit portion is 50\%. The course of $\alpha$
as a function of $\beta$ is plotted in Fig.~\ref{figa01}.

The temperature distribution calculated is this way results from the
assumption that the conductivities remains constant during the
iteration of eq.~\ref{eqa04}. This is not the case as $\kappa$ is
strongly temperature dependent. Therefore, after the new temperature
distribution is calculated, the conduction coefficients
have to be updated for this new temperatures and eq.~\ref{eqa04} has
to be solved again. This procedure is repeated until the error between
the new temperatures and the old ones is less then $10^{-4}$.

We have tested the method described above through the computation of
problems with known solutions. Two test cases were run, one
with a constant $\kappa$ and one with $\kappa \propto T^m$ with $m =
5/2$ for which a self similar solution exists (Zel'dovich et
al.~\cite{zr67}). The quality of the results for both test cases are
similar so only the results of the $\kappa =$const.-case are
presented here. 

In this test the plasma is assumed permanently static and with a
uniform and ``frozen'' density distribution with a particle density $n
= 6.6 \times 10^{-4}$ cm$^{-3}$. We have considered the propagation of
a spherical pure conduction front in a uniform medium. For this
problem an analytic solution for the temperature distribution as
function of time is available:
\begin{eqnarray}{\label{eqa09}}
T(r,t) = \left( \frac{ Q } {4 \, \pi \, \kappa \, t} \right)^{1.5} \,
\exp \left[ \frac{ - r^2 }{4 \, Q \, t} \right] \quad .
\end{eqnarray}
\begin{figure*}[h]
\psfig{figure=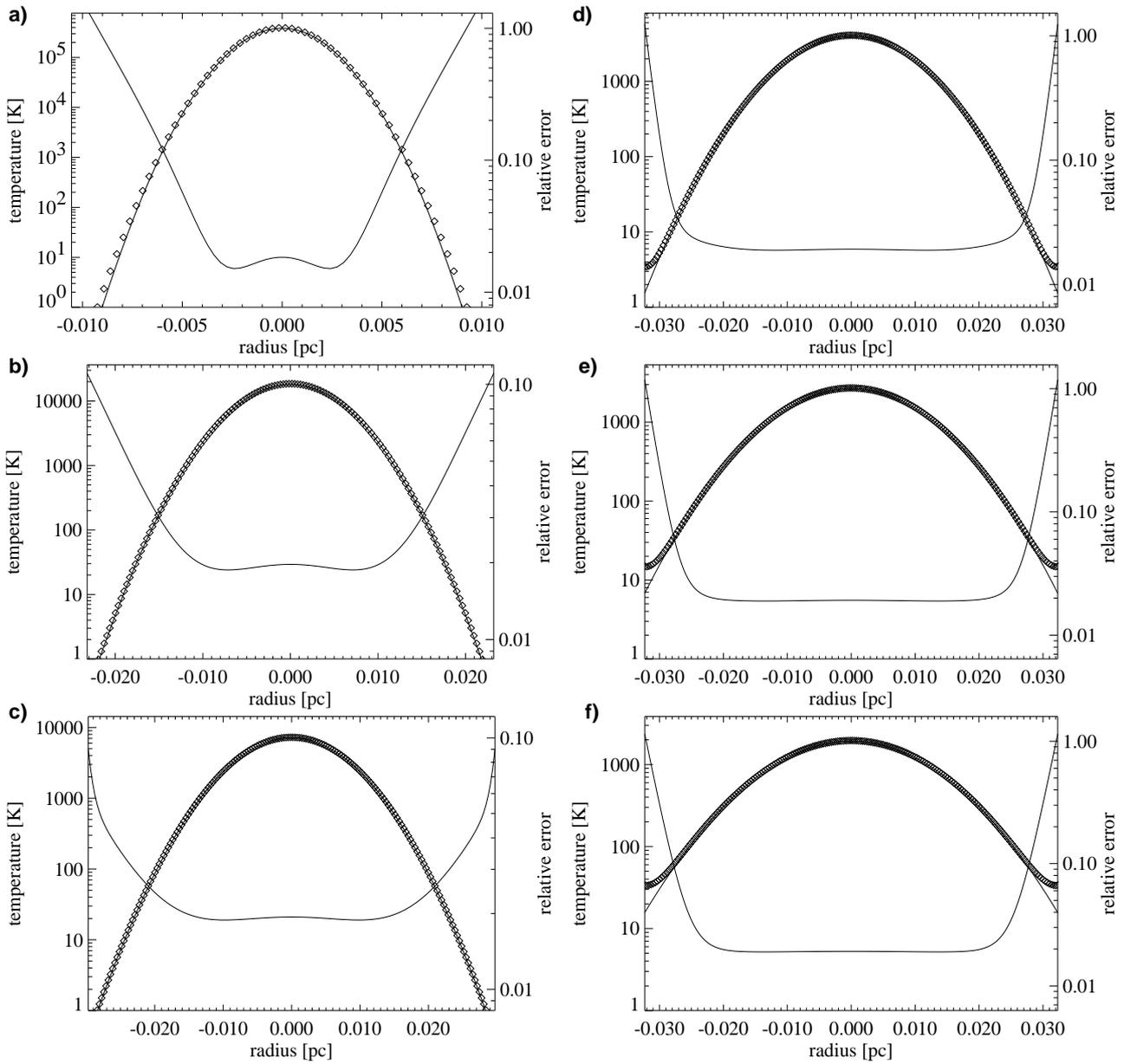,width=17cm}
\caption{\label{figa02}Propagation of the conduction front after 1.0s
(a), 6.0s (b), 11s (c), 16s (d), 21s (e) and 26s (f). Comparison
of the numerical results (diamonds) with the analytical solution
(solid line). Additionally, the relative error between both results is
plotted (convex line).
}
\end{figure*}
We have calculated the numerical solution in a cylindrical coordinate
system, taking as initial profile the analytical solution at $t =
0.5$ s, for $Q = 10^{54}$ K cm$^{-3}$ and $\kappa = 1.36 \times 10^{12}$
what corresponds to a conductivity of a plasma at a temperature of
$5.6 \times 10^6$ K. The grid is $200 \times 100$ cells with $-0.032 < z
< 0.032$ pc and $0 < r < 0.032$ pc. The resulting propagation, compared
with the analytical solution, and the relative error is shown in
Fig.~\ref{figa02} for different times. 
The relative error within the hottest region of the plasma amounts to
about 2-3\%. Only at the edge of the conduction front where the
temperature is some orders of magnitudes lower than in the inner
regions the error is slightly increased. In Fig.~\ref{figa02}d-f the
influence of the grid boundary is visible. Here the temperature
gradient is forced to be zero. With these studies we could verify the
validity and accuracy of the code. 

\end{appendix}

\end{document}